\documentclass[twocolumn,10pt]{IEEEtran}

\usepackage{multirow}
\usepackage[pdftex]{graphicx}
\usepackage{amsfonts,amsmath,amssymb,amsthm,epsfig,subfigure}
\usepackage{epsfig}

\newcommand{\be}{\begin{eqnarray}}
\newcommand{\ee}{\end{eqnarray}}

\def\eg{{\em e.g.},~}
\def\ie{{\em i.e.},~}

\newcommand{\ben}{\begin{enumerate}}
\newcommand{\een}{\end{enumerate}}
\newcommand{\beq}{\begin{equation}}
\newcommand{\eeq}{\end{equation}}
\newcommand{\beqa}{\begin{eqnarray*}}
\newcommand{\eeqa}{\end{eqnarray*}}
\newcommand{\bit}{\begin{itemize}}
\newcommand{\eit}{\end{itemize}}
\newcommand{\bt}{\begin{tabular}{c}}
\newcommand{\btt}{\begin{tabular}}
\newcommand{\et}{\end{tabular}}

\newtheorem{corollary}{Corollary}

\newtheorem{theorem}{Theorem}
\renewcommand{\baselinestretch}{0.978}
\begin{document}
\bibliographystyle{plain}

\title{Side-payment profitability 
under convex demand-response
modeling congestion-sensitive applications\thanks{This
research was supported by NSF CNS grant 1116626.}}

\author{G. Kesidis\\
CS\&E and EE Depts\\
Pennsylvania State University\\
University Park, PA, 16802\\
kesidis@engr.psu.edu\\
}

\maketitle

\begin{abstract}
This paper is concerned with the issue of side payments
between content providers (CPs) and Internet service 
(access bandwidth) providers (ISPs) 
in an Internet that is potentially not neutral. We herein 
generalize past results modeling the ISP and 
CP interaction as a noncooperative game in two directions. 
We  consider different
demand response models (price sensitivities) for different provider types
in order to explore when side payments are profitable to the
ISP.  
Also, we consider convex (non-linear) demand response to model 
demand 
triggered by traffic which is sensitive to access bandwidth congestion,
particularly delay-sensitive interactive real-time applications.
\end{abstract}

\section{Introduction}

Network neutrality continues to be debated as
its core economic issues as described in, \eg \cite{HW}, have not been resolved. 
The debate concerns all participants in the enormous and 
growing Internet economy:
Internet service (access) providers (ISPs), content
providers (CPs, including providers of cloud computing services),
end-user  consumers, and government regulators.

\begin{figure}[ht]
\begin{center}
\includegraphics[width=3.25in]{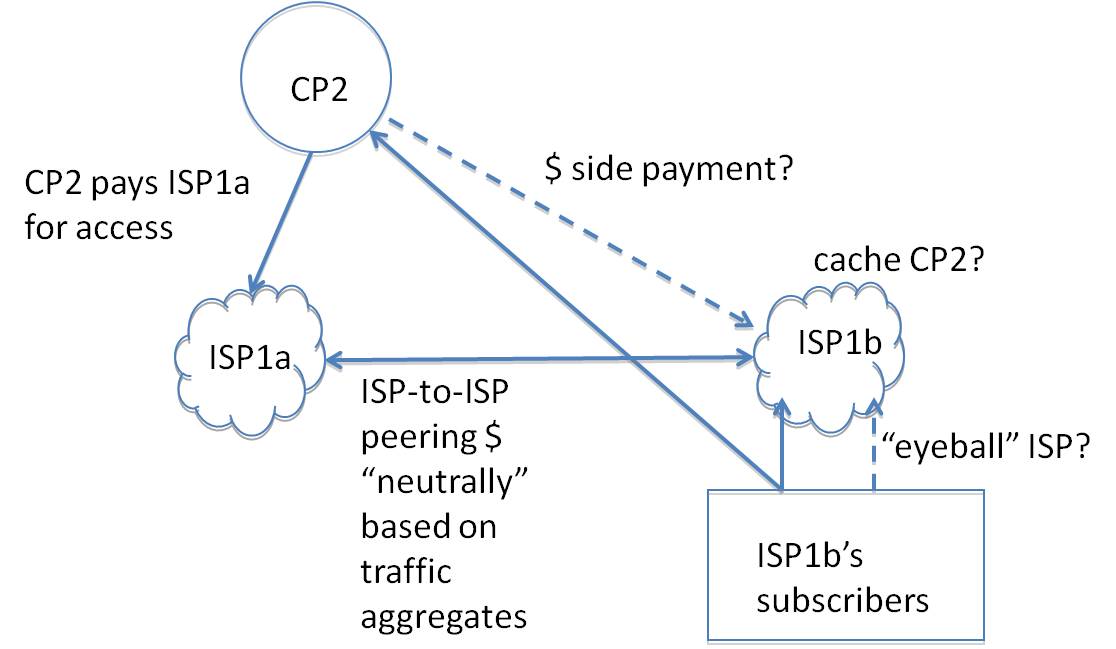}
\caption{Side payments issue}\label{sp-fig}
\end{center}
\end{figure}

In Figure \ref{sp-fig}, the subscribers of ISP1b 
directly pay ISP1b and content provider CP2.
CP2's Internet service provider, ISP1a, is different from ISP1b. 
The financial arrangements at the peering points
where the two ISPs exchange traffic are
typically ``neutral" in the sense that
they are based on traffic aggregates and not
on the (application) type of traffic\footnote{Note that
differentiated service (diffserv) among application types is ``neutral"
if requested by end-users,
whereas application diffserv implemented unilaterally by
an ISP is not application neutral
(whether implemented at ISP peering points or at end-user access points, and
even if for altruistic purposes, \eg to give more bandwidth
for putative interactive real-time applications).}
\cite{ReArch10,IFIP11} or its origin. 
Another question of neutrality is whether 
CP2 also pays ISP1b for access to its subscribers.
In the absence of such a ``side payment",
ISP1b might be disincentivized to locally cache ``remote" content
providers like CP2,
potentially gaining revenue at the ISP-to-ISP
peering points at the expense of additional
downloading delays experienced by ISP1b's subscribers.
In the following, we will not consider alliances
between independent CPs and ISPs (whose revenues may need to be dispersed
by, \eg  the use of Shapley values \cite{shap1,shap2}),
but we will 
consider peered ``eyeball" ISPs which also earns revenue as a content provider.

Also, we will not consider the role of advertising revenue 
\cite{walrand} in
Figure \ref{sp-fig}. Indeed, instead of directly subscribed
revenues, CPs may earn advertising revenue,
\eg \cite{IFIP11} 
(obviously, the subscribers ultimately pay by buying
the advertised merchandise or services). 
Note that ISPs have pointed out that
end-users do not necessarily ``request" of the
CPs the advertising
which the CPs insert into content, and have argued for sharing
of advertising revenue either through side payments
from CPs or for the right to insert their own ads into
downloaded content (the arrangements among the different parties
of broadcast television service may be analogous
to the latter scenario).

In the following, we do not extend the model of Figure \ref{sp-fig}
to study content exchanged peer-to-peer, which is expected to employ
only best-effort (\ie not usage-priced) service and, in practice, is
often in violation of copyright.
The enormous traffic volume associated with p2p activity was
one of the original motivators of the neutrality debate \cite{Anderson09}, and
issues of piracy continue to be developed, \eg \cite{Sisario11}.
Multiple types of content were previously modeled in, \eg
\cite{ReArch10}.

Finally, we do not consider user migration among multiple ISPs of the same type as in, \eg \cite{ReArch10}.

In this paper, we consider
the simpler model of side
payments depicted in Figure \ref{sp-fig}
on an end-user platform 
(\cite{hande,walrand} 
developed games involving end-users and content providers
on an ISP platform).
We review the case of a noncooperative
ISP-CP game based on a communal demand response for
providers of both types in Section \ref{sec-communal}.
The case of different, though coupled, demand responses
for each type of provider
is given in Section \ref{sec-diff} to explore the issue of 
profitability of side payments.
In Section \ref{sec-qos}, we consider
a convex demand-response corresponding
to acceleration in the demand for reserved bandwidth
by delay-sensitive (particularly interactive real-time)
applications once access bandwidth is congested; the
case of a communal demand response is taken here to simplify
analysis.
In Section \ref{smooth-convex}, we consider an example
of  a differentiable, convex demand-response for
side-payments\footnote{See \cite{Chiang11} for a recent
reference on side payments.}  between two {\em different} types of providers.
In an appendix,
we  also consider a different model involving two different eyeball ISPs 
connected at peering point(s) where revenue is generated corresponding
to net traffic transmitted.
Finally, we conclude in Section \ref{sec-concl} with a
short summary.

\section{Background: Communal demand response of an end-user 
platform}\label{sec-communal}
Suppose there are two providers, 
one content (CP indexed 2) and the other access (ISP
indexed 1).
Also, the {\em communal} consumer demand 
\cite{Economides} is assumed to be
\beqa
D & = & D_{\max}-d (p_1 + p_2),
\eeqa
where $d$ is demand sensitivity to the price,
and $D_{\max}>0$ is the demand at zero usage based price\footnote{Note
that ISPs  are continuing to depart from pure flat-rate pricing
(based on access bandwidth) for unlimited monthly volume,
\eg \cite{Shin06,ATT11}.}. 
The revenue of the ISP is
\be\label{content-based-rev1}
U_1 & = &  (p_1 + p_s)D,
\ee
where $p_s$ is the side payment from content to access provider.
Similarly, the revenue of the CP is
\be\label{content-based-rev2}
U_2 & = &  (p_2 - p_s)D.
\ee
Consider a noncooperative game played by the CP and ISP adjusting their
prices, respectively $p_2$ and $p_1$, to maximize
their respective revenues, with all
other parameters fixed\footnote{In particular, the fixed side-payment $p_s$ 
is here assumed regulated. Note that the utilities are linear
functions of $p_s$ so that if $p_s$ were under the control of one
of the players, $p_s$ would simply be set at an extremal value.}.

The question whether the ISP will profit from side payments is
not trivial because the side payment will naturally cause 
the CP to raise its prices thus lowering overall demand.
The short proof of following simple 
result follows from \cite{TelSys11,ReArch10} and
is included here in order to subsequently relate it to the
principal results of this paper.
We assume that the ISP will decide that it profits from side payments
only if this profit is strictly increasing in $p_s$ at $p_s=0$.

\begin{theorem}\label{old-thm}
At the interior Nash equilibrium\footnote{In this paper, we do not 
consider boundary Nash equilibria, where at least one player is 
selecting an extremal value for one of their control parameters,
often resulting in that player essentially opting out of the game,
or maximally profiting from it at the expense of the other player.},
requiring
\be\label{p_s-assumption}
| p_s| & <&  \frac{D_{\max}}{3d},
\ee
the ISP does not profit by the 
introduction of side payments.
\end{theorem}

\noindent
Note that the sign of $p_s$ may be negative, \ie the side payment
is from ISP to CP (\ie remuneration for content instead of access
bandwidth).

\noindent
\IEEEproof
The conditions for Nash equilibrium are:
\beqa
0  & = &   \partial U_1/\partial p_1 = D_{\max}-d(2p_1 + p_2 + p_s),\\
0  & = &   \partial U_2/\partial p_2 = D_{\max}-d(p_1 + 2p_2 - p_s).
\eeqa
So, the Nash equilibrium plays are
\beqa
p_1^* & = &  \frac{D_{\max}}{3d}-p_s,\\
p_2^* & = &  \frac{D_{\max}}{3d}+p_s.
\eeqa
Based on the previous two displays, we can see the requirement
of  (\ref{p_s-assumption}) for an interior Nash equilibrium
({\em i.e.}, $p_1^*,p_2^* >0$ and $D^*>0$).

The ISP's revenue at this Nash equilibrium is
\beqa
U_1^*   & = &   
\frac{D_{\max}^2}{9d}.
\eeqa
So, $U_1^*$ does not depend on $p_s$, \ie it is
constant, so not strictly increasing at $p_s=0$.
\qed

~\\
In \cite{ReArch10}, we showed that the
the ISP may actually experience a reduction in revenue/utility
with the introduction of side payments, using
a communal demand model that had different demand-price sensitivity
parameters $d$ per provider type and also multiple providers of each type.
Such a model was also considered in \cite{IFIP11}.
The following section fleshes this out for a special case.

\section{Different demand response depending on provider type}\label{sec-diff}

Several other previous works have used a communal
demand function for both content and access bandwidth even though
these are two different 
commodities with different dimensions. 
This said,  demand for
content and access bandwidth may be coupled in that pricing of one
will affect the other.

So, consider two {\em different}, 
coupled demand responses
for providers of type $k\in\{1,2\}$:
\be
D_1 & = & D_{\max,1} -d_{1}(p_1 +p_2),
\label{diff-dem-resp}\\
D_2 & = & D_{\max,2} -d_{2}(p_1 + p_2),
\nonumber
\ee
where we have maintained the same 
(a single) price sensitivity $d_k$ for each
type of demand $D_k$ for simplicity.
For example, the ISP's demand could be in terms of 
Mbps$=[D_1]$  while the, \eg  music CP's demand could be in terms of
songs/s$=[D_1]$. The point is that a single song's size could
vary from say $3-15$MB so, though correlated, the CP and ISP demands
are different.

\subsection{Side payments as a function of consumed access bandwidth}
Suppose the side payments {\em based on consumed access bandwidth}
 are factored as follows:
\be
U_1   =  (p_1 + p_s) D_1  & \mbox{and} & 
U_2   =  p_2 D_2  - p_s D_1. 
\label{bandwidth-based-sp}
\ee
Again, in this setting, we wish to find conditions on the 
of demand sensitivities $d_{i}$ and usage-based side payment price $p_s$
such that side payments are of value  to the ISP.

\begin{theorem}\label{bandwidth-based-sp-thm}
The introduction of  bandwidth based side payments (\ref{bandwidth-based-sp})
will profit the ISP if and only if
either
\be\label{profit-cond1}
\min\{2 \frac{D_{\max,2}}{D_{\max,1}},~1\} & > & \frac{d_2}{d_1}  
\ee
or 
\be\label{profit-cond2}
\max\{2 \frac{D_{\max,2}}{D_{\max,1}},~1\} & < & \frac{d_2}{d_1}.
\ee

\end{theorem}

\noindent
\IEEEproof
Let $p=p_1+p_2$ and $\delta_k = D_{\max,k}/d_k$.
After solving the first order
conditions $\partial U_k/\partial p_k = 0$, $k\in\{1,2\},$
we get that at Nash equilibrium,
\beqa
p^* & = &  \frac{1}{3}(\delta_1-p_s + \delta_2+\frac{d_1}{d_2}p_s),\\
p_1^* + p_s & = & \delta_1-p^*.
\eeqa
Thus, at Nash equilibrium,
\beqa
U_1^* & = & \frac{d_1}{9}
(2\delta_1-\delta_2+ (1-\frac{d_1}{d_2})p_s)^2.
\eeqa
So,
\be
\frac{\partial U_1^*}{\partial p_s} & = & \frac{2d_1}{9}(1-\frac{d_1}{d_2})
(2\delta_1-\delta_2+ (1-\frac{d_1}{d_2})p_s).
\label{U-diff}
\ee
This quantity is positive at $p_s=0$ if and only if
(\ref{profit-cond1}) or (\ref{profit-cond2}) hold.
\qed

~\\
So,  in this setting, positive side payments are not necessarily
profitable to the ISP (player 1).

\subsection{Discussion}\label{discussion-sec}

Potentially different 
sensitivities to {\em each} price  for each
type of demand were considered in 
\cite{kesidis11},   
\ie $D_k = D_{\max,k} -d_{k,k}p_k - d_{k,3-k}p_{3-k}$ for $k\in\{1,2\}$,
as well as competition among multiple providers,
thus extending the previous two theorems. 
 There it was found that
the ISP may not profit by the introduction of positive
side payments, depending on the parameters in play,
but numerically this was only the case when there was only one
provider of each type.

More generally,  $U_1^*$ is a complex nonlinear function of $p_s$
where it is possible that $U_1^*(p_s)$ might be greater than
$U_1^*(0)$ for some sufficiently large $p_s$ notwithstanding
$\tfrac{\partial U_1^*}{\partial p_s}(0)<0$. We expect 
that side payments will be simply rejected by ISPs in the event that 
$\tfrac{\partial U_1^*}{\partial p_s}(0)<0$.

\subsection{Discussion: Side payments as a function of consumed content}

Instead of (\ref{bandwidth-based-sp}),
suppose side payments are based on content and
factored as follows:
\beqa
U_1   =   p_1 D_1 +  p_s D_2, & &  
U_2   =   (p_2 - p_s) D_2.
\eeqa
Because total revenue $U_1+U_2$ is a constant function  of $p_s$,
the ISP's profit  is at the CP's expense, 
\ie  $\partial U_1^* /\partial p_s >0$ if and only if
 $\partial U_2^* /\partial p_s <0$. So, the latter condition is
similar to that of Theorem \ref{bandwidth-based-sp-thm} with
the indices swapped and the resulting modified conditions (\ref{profit-cond1})
or (\ref{profit-cond2}) are for profitability of the content provider.

\begin{corollary}
The introduction of content based side payments
will be  profitable to the content provider 
(\ie a loss for the ISP) if and only if
\beqa
\min\{2 \frac{D_{\max,1}}{D_{\max,2}},~1\} & > & \frac{d_1}{d_2}  
\eeqa
or 
\beqa
\max\{2 \frac{D_{\max,1}}{D_{\max,2}},~1\} & < & \frac{d_1}{d_2}.
\eeqa
\end{corollary}

\section{Access bandwidth congestion affecting demand 
response}\label{sec-qos}

For simplicity,  in this section we return
to the communal demand response scenario  of
(\ref{content-based-rev1}) and (\ref{content-based-rev2}).
In the following section, we will consider an example of
different demand responses for the problem context of this section.

\begin{figure}[ht]
\begin{center}
\includegraphics[width=3.25in]{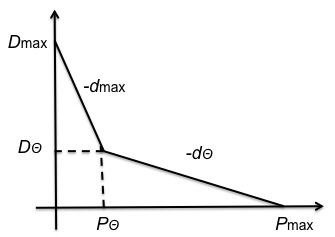}
\caption{Convex, piecewise linear demand-response}\label{pwl-fig}
\end{center}
\end{figure}

Suppose that there are two broad classes of applications,
one of which is significantly sensitive to congestion of access bandwidth,
\eg delay-sensitive interactive real-time applications.
Assume that applications of the other, best-effort type are
unlikely to engage in usage based-pricing for access bandwidth.
As pricing reduces,
the demand for access-bandwidth reservation increases, 
so causing additional congestion so that
best-effort service will be increasingly inadequate for
congestion-sensitive applications. Therefore,
the demand for usage-priced 
access-bandwidth reservation may {\em accelerate} with
reduced price.  More specifically,
say there is positive threshold 
\beqa
D_{\theta} & < & D_{\max}
\eeqa
such that overall demand sensitivity to price is greater 
when $D\geq D_{\theta}$ than when $D<D_{\theta}$. That is, for 
\beqa
d_{\max} & > & d_{\theta},
\eeqa
a {\em convex}, piecewise linear 
model for access bandwidth would be
\be\label{pwl-demand}
D(p)  & = &  
\max\{ D_{\max} - d_{\max} p,~ \hat{D}_{\theta} - d_{\theta} p\},
\ee
where
\beqa
\hat{D}_{\theta} & = &  D_{\theta} + (D_{\max} - D_{\theta}) d_{\theta}/d_{\max},\\
p_{\theta} & = &  (D_{\max} - D_{\theta})/d_m,\\
p_{\max} & = & \hat{D}_{\theta}/d_\theta ~=~
p_{\theta}+ D_{\theta}/d_{\theta},
\eeqa
so that $D(p_{\theta})=D_{\theta}$, see Figure \ref{pwl-fig}.

So, in this model, in the price range 
$[p_{\theta},p_{\max}]$
 (demand range $[0,D_{\theta}]$ equivalently) 
corresponds to low demand sensitivity 
to price, $d_{\theta}$.
The pricing range
$[0,p_{\theta}]$ (demand range $[D_{\theta},D_{\max}]$), 
when delay-sensitive applications typically need to adopt 
usage-priced (reserved or priority) access-bandwidth service,
corresponds to higher demand sensitivity to 
price, $d_{\max}$.

\subsubsection{Interior Nash equilibrium points}

Let $p:=p_1+p_2$ so that
\beqa
\frac{\partial U_1}{\partial p_1} & = & \left\{\begin{array}{ll}
D_{\max} -d_{\max}(p+p_1+p_s) & \mbox{if } p>p_{\theta}\\
D_{\theta} -d_{\theta}(p+p_1+p_s) & \mbox{if } p<p_{\theta}
\end{array}\right.\\
\frac{\partial U_2}{\partial p_2} & = & \left\{\begin{array}{ll}
D_{\max} -d_{\max}(2p-p_1-p_s) & \mbox{if } p>p_{\theta}\\
D_{\theta} -d_{\theta}(2p-p_1-p_s) & \mbox{if } p<p_{\theta}
\end{array}\right.
\eeqa
We take three cases for an interior
Nash equilibrium point (NEP) $(p_1^*,p_2^*)$ based on
comparing $p^*:= p_1^*+p_2^*$ to $p_{\theta}$.

~\\
{\bf Case $p^*>p_{\theta}$:}  
Adding the first order necessary conditions here gives
$p^*= 2D_{\max}/(3d_{\max})$. Checking ``consistency" with
the case condition requires
$p^*=2D_{\max}/(3d_{\max}) > p_{\theta}$, \ie if
\be\label{case1-cond}
3D_{\theta} ~ > ~  D_{\max}
& \mbox{and} & 
|p_s|~<~p^*/2,
\ee
then an interior NEP is $p_1^* =  p^*/2 - p_s$   and
 $p_2^* =  p^*/2 + p_s$.

~\\
{\bf Case $p^*<p_{\theta}$:}  Similarly, if 
\small
\be\label{case2-cond}
p^*=\frac{2D_{\theta}}{3d_{\theta}} ~ < ~ \frac{D_{\max}- D_{\theta}}{d_{\max}}
=p_{\theta}
& \mbox{and} & 
|p_s|~<~p^*/2,
\ee
\normalsize
then an interior NEP is 
$p_1^* =  p^*/2 - p_s$   and
 $p_2^* =  p^*/2 + p_s$.

~\\
{\bf Case $p^*=p_{\theta}$.} Here, for an interior NEP,
$\tfrac{\partial U_k}{\partial p_k}|_{p_{\theta} -}>0$
and 
$\tfrac{\partial U_k}{\partial p_k}|_{p_{\theta} +}<0$
 for $k=1,2$.
That is,  we have the following constraints on $p_1+p_s$:
\beqa
D_{\max} -d_{\max}(p_{\theta} +p_1^*+p_s) & < & 0 \\
D_{\theta} -d_{\theta}(p_{\theta} +p_1^*+p_s) & > & 0 \\
D_{\max} -d_{\max}(2p_{\theta} -p_1^*-p_s) & < & 0 \\
D_{\theta} -d_{\theta}(2p_{\theta} -p_1^*-p_s) & > & 0 
\eeqa
Equivalently,  for an interior NEP in this case
\be
\lefteqn{\max\{\frac{D_{\theta}}{d_{\max}}-p_s,~2p_{\theta}-\frac{D_{\theta}}{d_{\theta}}-p_s,~0\} 
 ~< ~ p_1^* } & & \nonumber\\
 & <  &
\min\{\frac{D_{\max}-2D_{\theta}}{d_{\max}}-p_s,~\frac{D_{\theta}}{d_{\theta}}-p_{\theta} -p_s,~p_{\theta}\},
\label{case3-cond}
\ee
where we have also used 
$0< p_1^* \leq p^*=p_{\theta}$ (the equality for this case).  

~\\
For (\ref{case1-cond}) and (\ref{case2-cond}), the issue
of profitability of side payments for the ISP (player 1)
is just as in Theorem \ref{old-thm}.

\subsection{Numerical examples for piecewise-linear demand response}

Consider the numerical example
where $D_{\max}= 2.5 \cdot D_{\theta}=1$,
$d_{\max}=5d_{\theta}=1$ (both set equal to
1 without loss of generality).
So, $p_{\theta}=(1-0.4)/1=0.6$.
Condition (\ref{case1-cond}) clearly holds, giving
the corresponding NEP if $|p_s|<p^*/2$, but  conditions
(\ref{case2-cond}) and (\ref{case3-cond}) cannot hold.

For a second numerical example, 
take $D_{\max}= 6 D_{\theta}=1$ and
$d_{\max}=6d_{\theta}=1$. 
In this example,
(\ref{case1-cond}) and (\ref{case3-cond}) cannot hold.
Here, $p^*=2/3 < 5/6 = p_{\theta}$, so 
(\ref{case2-cond}) does hold giving the
corresponding NEP if $|p_s|<p^*/2$.

\begin{figure}[ht]
\begin{center}
\includegraphics[width=3.25in]{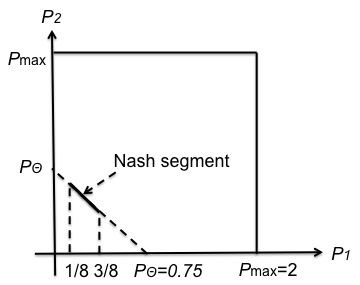}
\caption{Nash equilibrium line segment when
$D_{\max}= 4 D_{\theta}=1$,
$d_{\max}=5d_{\theta}=1$, and $p_s=1/8$.}\label{nash-segment}
\end{center}
\end{figure}

\begin{figure}[ht]
\begin{center}
\includegraphics[width=3.25in]{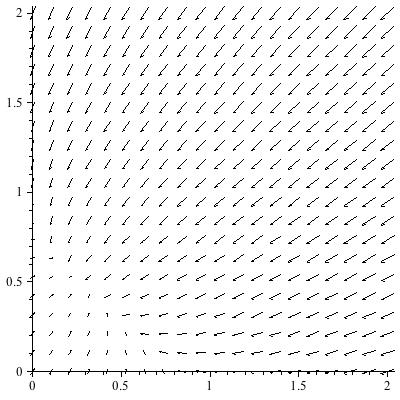}
\caption{Vector field of the game with 
$D_{\max}= 4 D_{\theta}=1$,
$d_{\max}=5d_{\theta}=1$, and $p_s=1/8$.}\label{nash-quiver}
\end{center}
\end{figure}

For a final numerical example, 
take $D_{\max}= 4 D_{\theta}=1$ and
$d_{\max}=5d_{\theta}=1$.
Here, $p_{\theta} = (1-0.25)/1 =0.75$ and
$p_{\max}=p_{\theta}+D_{\theta}/d_{\theta} = 2$.
In this example,
(\ref{case1-cond}) and (\ref{case2-cond}) do not hold;
but  when $p_s< 0.5$,
(\ref{case3-cond}) reduces to
\be\label{Nash-seg}
0.25-p_s & < ~ p_1^* ~ < & 0.5-p_s.
\ee
That is, there is a Nash equilibrium line
segment $(p_1^*,p_2^*)$ such that $p_1^*$
satisfies (\ref{Nash-seg}) and 
$p_2^*=p_{\theta}-p_1^* = 0.75-p_1^*$. See
Figure \ref{nash-segment} for the case whter $p_s=1/8$.
In Figure \ref{nash-quiver}, we depict
the vector field 
$(\tfrac{\partial U_1}{\partial p_1}, 
\tfrac{\partial U_2}{\partial p_2})$ as a function of
$(p_1,p_2)$,
corresponding to ``better-response"
time-differential, synchronous-play Jacobi dynamics \cite{Jin05,Shamma05},
\beqa
\frac{\mbox{d} p_k}{\mbox{d} t} & = & 
\frac{\partial U_k}{\partial p_k} - p_k,~~\mbox{for} ~k=1,2,
\eeqa
of this noncooperative game between ISP and CP.

\section{Differentiable, convex demand model}\label{smooth-convex}

We can also consider a convex, differentiable demand model
that can approximate (\ref{pwl-demand}), specifically
\be\label{convex-demand}
D(p) & = & D_{\max}(1-p/p_{\max})^{\alpha},
\ee
where $p=p_1+p_2$. Here, $\alpha\geq 1$ and 
 given $d_{\max}>d_{\theta}>0$ and $0<D_{\theta}<D_{\max}$,
$p_{\max}$ may be  found
using $D'(0)=-d_{\max}$ and 
$D'((D)^{-1}(D_{\theta}))=
 D'(p_{\theta})= -d_{\theta}$,

In this section, we first study the system described above for
this model of demand response. We then consider a different system
of two different ``eyeball" ISPs (\ie combined CPs and ISPs) with
an inter-provider service-level agreement (SLA) involving a revenue
based on net traffic flow across their peering point(s).

\subsection{Interior Nash equilibria and side-payments
under different demand-response for each type of provider}

\begin{theorem}
At interior Nash equilibrium for a strictly convex demand response,
\be\label{pstar-k}
p^*_1 =   p^*/2 - p_s &\mbox{and} & 
p^*_2 =   p^*/2 + p_s,
\ee
where $p^*=p^*_1+p^*_2$ solves 
\be\label{pstar-cond}
2D(p^*)+p^*D'(p^*) & = & 0.
\ee
and $|p_s|<p^*/2$.
\end{theorem}

\noindent
\IEEEproof
Again, (\ref{pstar-cond}) is the sum of the 
first-order conditions  for a 
NEP,
$\sum_k \partial U_k/\partial p_k =0$.
Conclude by substituting $p^*$ into each
first-order condition
to obtain (\ref{pstar-k}). 
\qed

~\\
For the example of (\ref{convex-demand}) with $\alpha > 1$,
\beqa
p^* &  = & \frac{2}{2+\alpha} p_{\max},\\
U_1^* & = & \frac{p^*}{2}D(p^*)~=~
 \frac{p_{\max}}{2+\alpha} \left(1-
 \frac{2}{2+\alpha} \right)^{\alpha}.
\eeqa

Again, note that under communal demand-response,
neither $p^*=p_1^* + p_2^*$ nor $U_1^*$ depend on $p_s$.
This is obviously not necessarily the case when the demand-response
for ISP and CP are different as in the previous section
and \cite{kesidis11}.  So,
now suppose for this example that the CP and ISP demands differ,
but  only in the $D_{\max}$ parameter, \ie 
\beqa
U_1 & =&  D_{\max,1}(1-(p_1+p_2)/p_{\max})^{\alpha}(p_1+p_s),\\
U_2 & = & D_{\max,2}(1-(p_1+p_2)/p_{\max})^{\alpha}p_2 \\
& & ~~-D_{\max,1}(1-(p_1+p_2)/p_{\max})^{\alpha}p_s.
\eeqa
For this example, at the interior Nash equilibrium 
\beqa
\frac{\partial U_1^*}{\partial p_s}(0) & = & 
\frac{D_{\max,1}(\tfrac{\alpha}{2+\alpha})^{\alpha} (D_{\max,2}-D_{\max,1})(1+\alpha)}{D_{\max,2}(2+\alpha)}.
\eeqa
So, the introduction of access-bandwidth based
side payments will be profitable to
the ISP, \ie
$\tfrac{\partial U_1^*}{\partial p_s}(0) >0$,
 if and only if 
\beqa
D_{\max,2} & > & D_{\max,1}.
\eeqa
A similar result can be likewise derived for content-based side-payments.


\section{Summary and Conclusions}\label{sec-concl}

In summary, we considered three different problems modeling
demand response in an access network game. The first was
to explore the profitability of side payments under (reasonably)
different demand response for ISP and CP. We found that
side-payments from the CP to the ISP 
were {\em not} unconditionally profitable
for the ISP.
The second problem analyzed a convex demand response
(taken to be communal to simplify analysis) that was
used to model the onset of demand for access bandwidth by 
delay-sensitive applications
once the access is sufficiently congested.  Note that the
former problem is one pertaining to the network neutrality
debate, while the latter problem setting
would be application neutral
if preferential treatment (service priority or bandwidth
reservations) of applications by the ISP is performed
at the behest of the consumer/user.



\section*{Appendix: Two different eyeball ISPs}

We now consider a game focusing
on two different eyeball ISPs, 
denoted  by subscripts  ``$a$" and ``$b$",
on a platform
of users and CPs, \ie the ISPs also serve
as CPs so no separate pricing by CPs is modeled.
 For $k,j\in\{a,b\}$,  
the demand for ISP $k$'s content is $D_k(p_j)$ when it is 
based on ISP $j$'s access-bandwidth price $p_j$. 
In the following, the same price $p_j$ will
be used by ISP $j$ irrespective of content source, \ie content
is neutrally priced in this sense.

Suppose there are peering points between these
two ISPs where net transit traffic flow in one direction
will correspond to net revenue for the (net) receiving
ISP at rate $p_t$  from the (net) transmitting ISP\footnote{For example,
France telecom charges $p_t=$\$3/Mbps, whereas pricing from the DSLAM
to core, \ie access bandwidth, for their {\em content providers} is \$40/Mbps.
This said, many existing peering agreements among non-transit ISPs have no transit pricing, \ie $p_t=0$.}.
See \cite{Dovrolis08,Feamster11} for recent studies of models
of transit pricing 
for a network involving a transit ISP between the content providers
and end-user ISPs.

Without caching,
transit traffic volume is obviously maximal and remote content
may be subject to additional delay
possibly increasing  demand (reducing demand sensitivity) for
usage-priced bandwidth reservations.
However, poorer delay performance
may instead reduce demand for remote content
or cause subscribers to change to ISPs that cache remote content.
Let  $\Phi_k$ be the fraction of  remote demand
that is {\em not} cached by ISP$k$.
By simply separately accounting for the demand by two different
user populations with similar content preferences, 
we take the utilities as:
\beqa
U_a(p_a) & = & D_a(p_a)p_a +[\Phi_b D_a(p_b)-\Phi_a D_b(p_a)]^+ p_t,\\
U_b(p_b) & = & D_b(p_b)p_b +[\Phi_a D_b(p_a)-\Phi_b D_a(p_b)]^+ p_t,
\eeqa
where $[x]^+:=\max\{x,0\}$ in the second (transit revenue) terms. So,
we have assumed
different ``upstream" congestion points for local and remote traffic  and
no revenue from cached (best-effort) traffic.
Note that $\Phi_k\leq 1$ will be chosen by ISP$k$ at its {\em minimal}
value, which we here assume to be strictly positive again
because an ISP that does not cache any remote content may lose subscribers,
or demand for remote content may be reduced,
owing to poor delay performance.
We will also assume that $p_t$ is fixed and,
by volume discount, $p_t < \min\{p_a,p_b\}$.

We study this system for an example of convex demand.
Again suppose, for $k\in\{a,b\}$,
\beqa
D_k(p) &= & D_{\max,k}(1-p/p_{\max})^{\alpha},
\eeqa
where the maximal price $p_{\max}>0$  and $\alpha\geq 1$
are also assumed to be common parameters for both ISPs.
Without loss of generality, assume the demand factor
\be\label{wlog-assumption}
\delta  & := &  D_{\max,b}/D_{\max,a} ~< 1,
\ee
\ie demand for ISP $a$'s content is generally higher
than that of ISP $b$.
Also define the caching factor
\be
\varphi  & := &  \Phi_{b}/\Phi_{a}.
\ee

~\\
{\bf Case $\Phi_b D_a(p_b^*) > \Phi_a D_b(p_a^*)$:}
The first order conditions for Nash equilibrium are
\beqa
 D_a'(p_a)p_a + D_a(p_a)- \Phi_b D_b'(p_a)p_t & = & 0\\
 D_b'(p_b)p_b + D_b(p_b) & = & 0.
\eeqa
The solution here is:
\beqa
p_b^*  ~= ~ \frac{1}{1+\alpha}p_{\max} & < & 
p_a^* ~ = ~ p_b^*
+\frac{\alpha\Phi_b\delta}{1+\alpha} p_t.
\eeqa
The requirement $p_t < p_b^*<p_a^*<p_{\max}$ gives 
the following condition on $p_t$ for
an interior equilibrium in this case:
\beqa
\frac{p_t}{p_{\max}} & < & \min\left\{\frac{1}{1+\alpha},~ 
\frac{1}{\Phi_b\delta}\right\}.
\eeqa
Adding the case condition, $\Phi_b D_a(p_b^*) > \Phi_a D_b(p_a^*)$,  gives
\be
 \frac{p_t}{p_{\max}} & < &  \min\left\{\frac{1}{1+\alpha}, ~
\frac{1-(\varphi/\delta)^{1/\alpha}}{\Phi_b\delta}\right\}.
\label{p_t-bound}
\ee
Note that we require $\varphi<\delta$ (caching factor is
less than demand factor), otherwise
this bound is $p_t<0$. In particular, this implies
that $\varphi<1$
because of  (\ref{wlog-assumption}).

~\\
{\bf Case $\Phi_b D_a(p_b^*) < \Phi_a D_b(p_a^*)$:}
Now,
\beqa
p_a^*  ~= ~ \frac{1}{1+\alpha} p_{\max}& < & 
p_b^* ~ = ~ 
p_a^*  +
\frac{\alpha \Phi_a/\delta }{1+\alpha}p_t.
\eeqa
Again,
for an interior Nash equilibrium,  
the case condition itself is an additional parametric constraint.
Here, these requirements lead to:
\beqa
\frac{p_t}{p_{\max}}& < & 
\min\left\{
\frac{1}{1+\alpha},~
\frac{1-(\delta/\varphi)^{1/\alpha}}{\Phi_a/\delta }
 \right\}.
\eeqa
So, here we require $\varphi>\delta$ so that $p_t>0$.

%
%

\end{document}